\documentclass[pre,floats,aps,superscriptaddress,showpacs]{revtex4}

\usepackage{graphicx}
\usepackage{amsmath}
\usepackage{version} 

\newtheorem{definition}{Definition}

\newcommand{\bv}{{\bf v}}

\newcommand{\bs}{\hat{\boldsymbol \sigma}}

\begin{document}
\title{The dynamical collision network in granular gases.}

\date{\today}

\author{J. Ignacio Alvarez-Hamelin}
\affiliation{CONICET and Facultad de Ingenier\'{\i}a,
Universidad de Buenos Aires, Paseo Col\'on 850, C1063ACV Buenos Aires,
Argentina}

\author{Andrea Puglisi}
\affiliation{Dipartimento di Fisica, Universit\`a La Sapienza, p.le Aldo
Moro 2, 00185 Roma, Italy}

\begin{abstract}
  We address the problem of recollisions in cooling granular
  gases. To this aim, we dynamically construct the interaction
  network in a granular gas, using the sequence of collisions
  collected in an event driven simulation of inelastic hard disks
  from time $0$ till time $t$. The network is decomposed into its
  $k$-core structure: particles in a core of index $k$ have collided
  at least $k$ times with other particles in the same core. The
  difference between cores $k+1$ and $k$ is the so-called $k$-shell,
  and the set of all shells is a complete and non-overlapping
  decomposition of the system. Because of energy dissipation, the gas
  cools down: its initial spatially homogeneous dynamics,
  characterized by the Haff law, i.e. a $t^{-2}$ energy decay, is
  unstable towards a strongly inhomogeneous phase with clusters and
  vortices, where energy decays as $t^{-1}$.  The clear transition
  between those two phases appears in the evolution of the $k$-shells
  structure in the collision network. In the homogeneous state the
  $k$-shell structure evolves as in a growing network with fixed
  number of vertices and randomly added links: the shell distribution
  is strongly peaked around the most populated shell, which has an
  index $k_{max} \sim 0.9 \langle d \rangle$ with $\langle d \rangle$
  the average number of collisions experienced by a particle. During
  the final non-homogeneous state a growing fraction of collisions is
  concentrated in small, almost closed, {\em communities} of
  particles: $k_{\max}$ is no more linear in $\langle d \rangle$ and
  the distribution of shells becomes extremely large developing a
  power-law tail $\sim k^{-3}$ for high shell indexes. We
  conclude proposing a simple algorithm to build a correlated random
  network that reproduces, with few essential ingredients, the whole
  observed phenomenology, including the $t^{-1}$ energy decay. It
  consists of two kinds of collisions/links: single random collisions
  with any other particle and long chains of recollisions with only
  previously encountered particles. The algorithm disregards the exact
  spatial arrangement of clusters, suggesting that the observed
  string-like structures are not essential to determine the statistics
  of recollisions and the energy decay.
 \end{abstract}

\pacs{45.70.-n,51.10.+y,89.75.Hc}

\maketitle

\section{Introduction}

The study of dilute granular materials~\cite{intro1}, as well as of
other many particles systems lacking an equilibrium description (glass
forming liquids in their non-equilibrium regime~\cite{intro2}, for
instance), requires the use of appropriate tools, models and paradigms
in order to simplify the picture and gain new insights. The
introduction of the free inelastic hard spheres model has been of
fundamental relevance for the rigorous foundation of granular kinetic
theory and hydrodynamics~\cite{introgas}. The model consists of a gas
of $N$ hard spheres in a volume $V$, with inelastic collisions
modeling the macroscopic nature of grains: part of the energy
of the relative motion between two colliding grains is irreversibly
dissipated into heat. Such simple generalization of the hard sphere
model leads to dramatic consequences: the gas performs a
non-stationary dynamics, cooling down and evolving toward a final
state of thermal death. Moreover, in large enough systems a
homogeneous granular gas is unstable toward the formation of vortices
and clusters of closely packed grains.

A satisfying description of the late dynamics of the gas, after the
onset of strong inhomogeneity, is still lacking.  Coarse-grained
descriptions in terms of hydrodynamic fields are made difficult by an
apparently general lack of separation between microscopic and
macroscopic scales~\cite{goldhirschchaos,cecco}. Velocity or density structure factors, usually
difficult to be measured because of rapid temporal evolution, indicate
a generic growth of structures but are insufficient to appreciate the
rich kinetic behavior at the level of small groups of
grains. Analogies with percolation transitions, Burgers models, or
$2D$ turbulence have been proposed, sometimes on the mere basis of
phenomenological similitudes.

In front of the openness of this problem, we propose a novel method of
analysis, borrowed to the graph's theory, the so-called $k$-core
analysis~\cite{Batagelj02,nacho,doro}: a $k$-core is the maximum
subgraph containing only nodes with at least $k$ links in the same
subgraph. Very recently, the theory of $k$-cores has proved to be
useful in defining a possible universality class for the dynamical
glass transition, the bootstrap (also called ``jamming'')
percolation~\cite{glasses}. In simple models of glass forming liquids,
at finite temperature and above a critical value of the tunable
average density, a jammed giant component in the $k$-core subset of
the contact network (defined in a specific way) percolates the system,
marking a breaking of ergodicity. Even if these works have been of
some inspiration to us, we stress that our study addresses a different
phenomenology and applies the analysis of graphs in a substantially
different way. Therefore we do not claim any analogy between the
dynamical breaking of the Homogeneous Cooling State of a granular gas
and the jamming percolation or any other proposed glass transition
scenario.

Our proposal is to dynamically construct a graph by linking two nodes
(particles) each time a collision happens: the resulting graph grows
with time and the average number of links per node is equal to two
times the average number of collisions per particle. The decomposition
of the interaction network into a hierarchy of cores of increasing
degree of self-interaction will prove to be a powerful tool to
distinguish between particles in the solid-like phase from those in
the fluid-like phase.  The sequence of collisions in the
non-homogeneous state is in fact dominated by repeated impacts among
grains trapped into sorts of self-sustained cages with very small
escape probabilities. Such cages can be thought as almost closed {\em
communities} in an interaction network. Strongly dissipating
collisions, on the other side, are mostly concentrated outside of
those cages, in the periphery of clusters, and involve particles which
still have a certain degree of mobility.

\begin{figure}[htbp]
{\includegraphics[width=16cm]{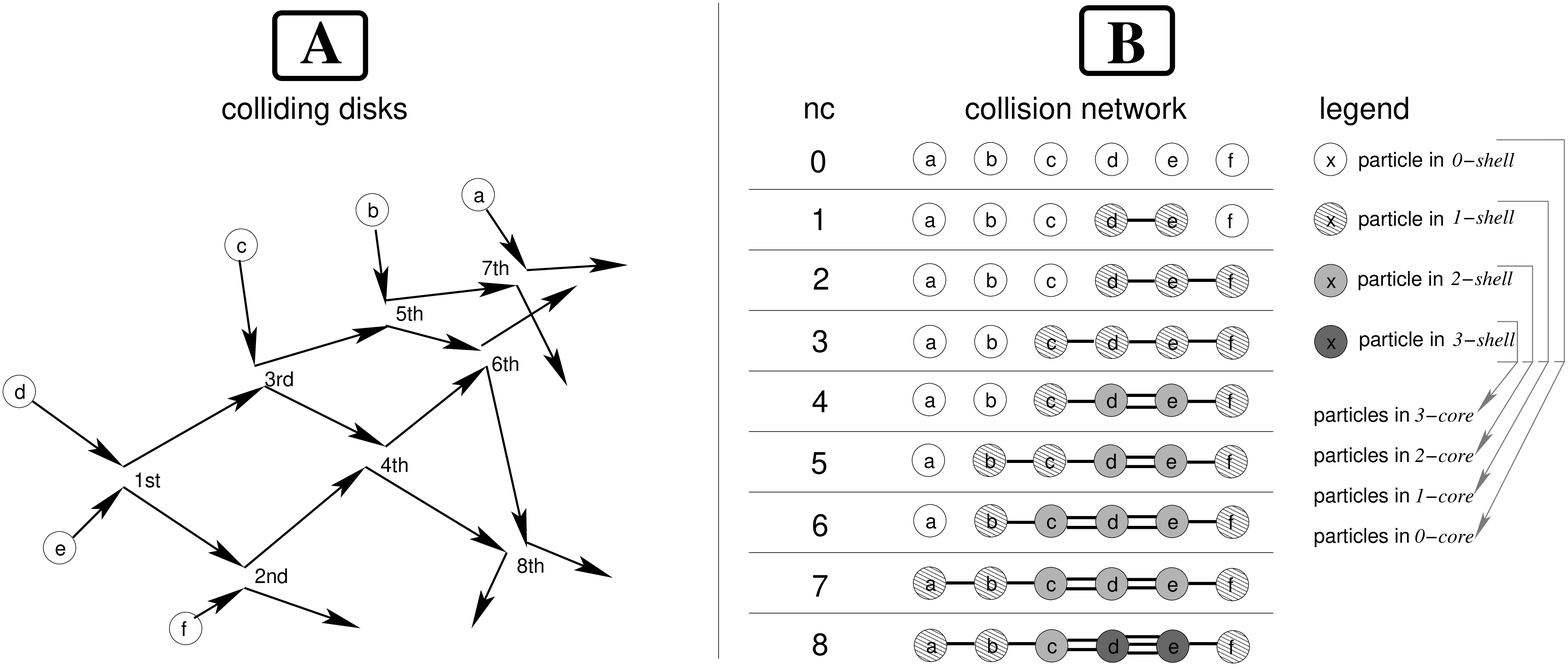}}
\caption{\label{fig:rings}Evolution of the network of collisions for a
system of 6 particles.  {\bf A}: the ordered sequence of
collisions. {\bf B}: collision network at each time step, where $nc$ is
the number of realized collisions. All particles are in core of index
$0$. Particles not appearing in higher cores are also in shell of
index $0$. In general a $k$-core has all the particles with shell
index $\geq k$.  }
\end{figure}

More specifically, the $k$-core decomposition of the collision network
appears as the natural characterization of the violation of Molecular
Chaos. In an ideal gas, i.e. a gas satisfying the Molecular Chaos
assumption, a grain never collides with another grain that is
connected (through collisions) to its own previous history. In terms
of graphs, true Molecular Chaos is equivalent to the following
statement: the collision network is a tree, and therefore all the
cores of index larger than $1$ are empty. The minimal mechanism of
Molecular Chaos breaking is that of {\em rings collisions}, which
appear as loops in the collision graph, i.e. the simplest structures
in cores of index $2$. An example of dynamical collision network, with
trees and loops, is given in Fig.~\ref{fig:rings}. Cores of high
degrees, i.e. systematic violations of Molecular Chaos, appear
naturally in any finite system, even with purely random collisions.
We will focus on the difference between effects on the $k$-core
decomposition due to finite-size and those really induced by
inelasticity. 

In section II a description of the model of inelastic hard disks and
of his known regimes will be given. The construction of the dynamical
collision network will be explained in section III. In section IV we
will show numerical results for the homogeneous state, while the
non-homogeneous state will be discussed in section V. In
section VI we discuss a model of random correlated collisions
that reproduces the main features observed in a non-homogeneous
granular gas. A discussion with conclusions and perspectives will
come in section VII.

\section{The granular gas of inelastic hard spheres}

One of the simplest models of granular gas is the gas of free
inelastic hard objects (rods, disks or spheres, depending on the
dimensionality) with constant coefficient of restitution. We restrict
the discussion to $D=2$ dimensions, considering an assembly of $N$
inelastic hard disks of diameter $1$ and mass $1$ a in a box of volume
$V=L\times L$ with periodic boundary conditions. Between collisions
each particle moves freely. When two particles $i$ and $j$ collide
their velocities change from $\bv_i,\bv_j$ to $\bv_i',\bv_j'$, using
the following rule:

\begin{equation} \label{eq:rule}
\bv_{i(j)}'=\bv_{i(j)}-\frac{1+\alpha}{2}[(\bv_{i(j)}-\bv_{j(i)})\cdot \bs]\bs,
\end{equation}
where $\alpha \in [0,1]$ is the restitution coefficient, while $\bs$
is the unity vector joining the centers of the two particles.

\begin{figure}[htbp]
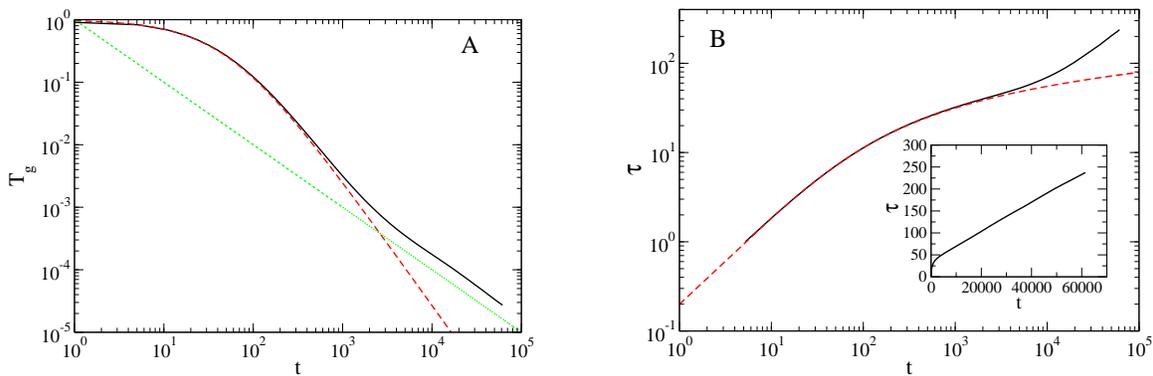

\includegraphics[clip=true,width=7cm]{energy} \hspace{1cm}
\includegraphics[clip=true,width=7cm]{ncoll}
\caption{\label{fig:global}{\bf (Color online)} {\bf A}: decay of
granular temperature $T_g$ vs. time $t$. The dashed red line marks the
Haff law, $T_g=1/(1+t/t_e)^2$ with $t_e \approx 52$, the dotted green
line is a $t^{-1}$ decay. {\bf B}: cumulated number of collisions per
particle $\tau$ vs. time $t$, the dashed red line shows the
logarithmic growth of collisions in the Haff phase, $\tau=a
\ln(1+t/t_e)$ with $a \approx 10$. In the inset a linear plot of
$\tau$ vs. $t$ is shown, focusing on the behavior at large times ($t
\gg 10^4$). The system is a gas of $N=10^6$ inelastic hard disks with
restitution coefficient $\alpha=0.9$ in a box of size $3163 \times
3163$ (units of a diameter) with periodic boundary
conditions. Averages over $10$ different initial conditions have been
performed. Units in the figures are arbitrary, based on simulation
parameters.}
\end{figure}

This model is widely used in the theoretical work on granular gases,
conveying usefulness and simplicity: it reproduces the main features
of real dilute granular experiments, allows fast event driven
simulations and is prone to analytical formalization, being at the
very basis of rigorous kinetic theories~\cite{introgas}. A vast
campaign of studies has been devoted, in recent years, to understand
the properties of this model during the so-called Homogeneous Cooling
State (HCS) and the linear departure from it. The system is prepared
in a random homogeneous configuration with positions uniformly
distributed in the box and velocities extracted from a Maxwellian with
temperature $T_g(0) \equiv \langle \bv_i^2(0)/2 \rangle=1$. A
thermalization phase follows, where the system evolves with elastic
collisions. Finally the real inelastic evolution begins: all time and
collision counters start from this moment. During evolution, the
granular temperature decreases since a fraction of kinetic energy is
lost after each collision. Nevertheless the initial stage of the
cooling remains homogeneous in space, with a decay of the granular
temperature given by the Haff law:
\begin{equation} \label{eq:haff}
T_g(t)=\frac{T_g(0)}{(1+t/t_e)^2}
\end{equation}
with $t_e=\tau_e/\gamma$, $\tau_e$ the mean free time between
collisions at the beginning of the evolution, and $\gamma=(1-\alpha^2)/4$.
During the HCS, the cumulated number of collisions per particle, $\tau(t)$,
grows logarithmically with time:
\begin{equation}
\tau(t) \propto \ln(1+t/t_e).
\end{equation}
In Fig.~\ref{fig:global} we show $T_g$ vs. $t$ (frame A) as well
as $\tau$ vs. $t$ (frame B).

\begin{figure}[htbp]
\parbox{5.5cm}{\includegraphics[clip=true,width=5cm]{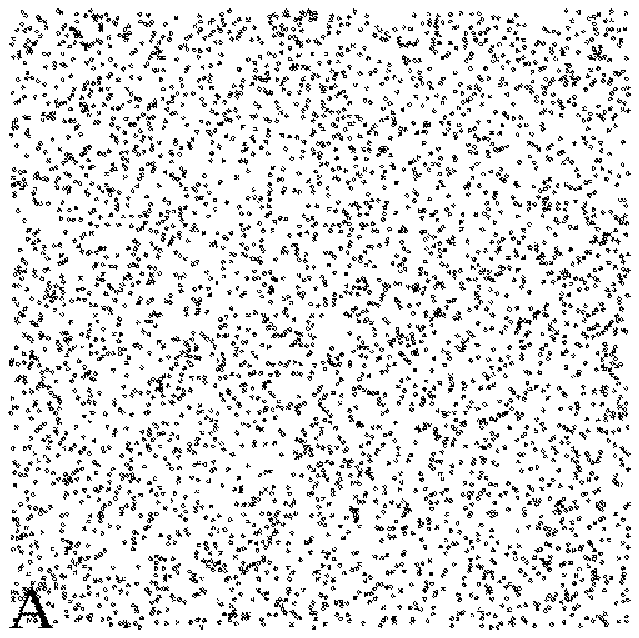}}
\parbox{5.5cm}{\includegraphics[clip=true,width=5cm]{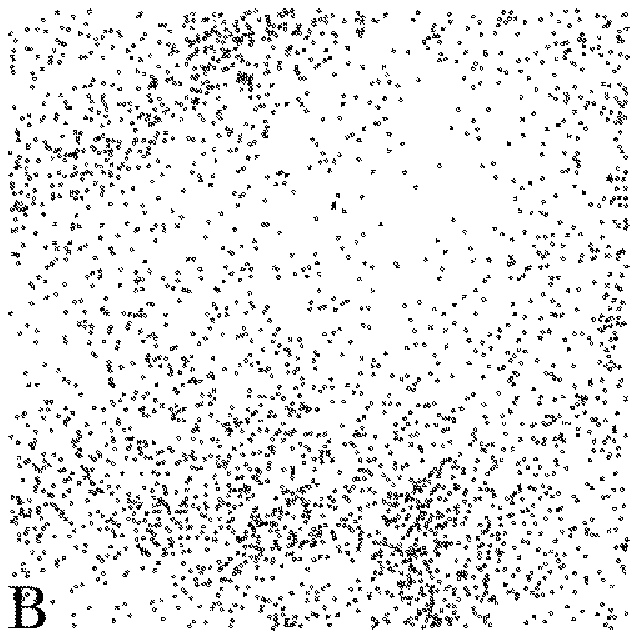}}
\parbox{5.5cm}{\includegraphics[clip=true,width=5cm]{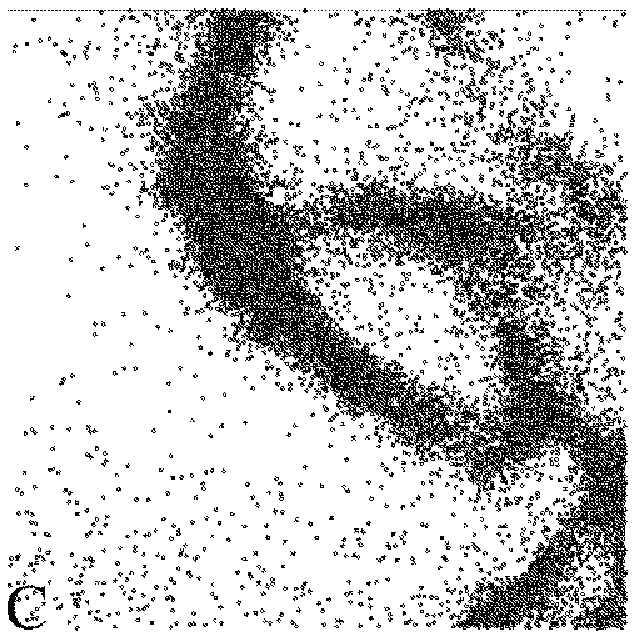}}
\caption{\label{fig:frames}Three snapshots of a small part of the gas
of $10^6$ inelastic hard disks with restitution coefficient
$\alpha=0.9$. In the snapshots is portrayed a square of surface $200
\times 200$, in units of a diameter, where the whole system is $3163
\times 3163$; therefore the pictures portray a fraction $0.004$ of the
total surface. {\bf A}: just after thermalization ($0$ collisions per particle (cpp)). {\bf B}:
after $100$ cpp (collisions per particle). {\bf C}: after $200$ cpp. }
\end{figure}

Simulations and linear stability analysis of macroscopic
(hydrodynamic-like) models~\cite{mcnamara,goldhirsch,ernst}, have
shown that the HCS is unstable when the size of the system $L$ is
larger than a critical size $L_c$. For systems large enough,
therefore, spatial homogeneity is eventually broken and the formation
of vortices and clusters is observed. A sign of this symmetry breaking
is the departure from the Haff law: the growth of structures reduces
the dissipation, reflected by a slower decay of $T_g$. The generally
accepted scenario is a two-stages process: initially the density
remains homogeneous while a macroscopic velocity field with vortical
structures emerges (shear instability); after a while the density
field begins to develop clusters (clustering instability). Both kinds
of structures seem to obey a coarsening dynamics with characteristic
lengths growing as $\sim t^{1/2}$~\cite{ernst}. In
Fig.~\ref{fig:frames} a sequence of snapshots of a granular gas
simulation illustrates the breakdown of the Homogeneous Cooling State
and the onset of a highly clusterized regime.

Recently the evidence of an asymptotic non-homogeneous cooling
dynamics has been shown in large ($N \sim 10^6$) event driven 2D and
3D simulations~\cite{bennaim,brito,isobe,luding_new}. The observation
of this ``final'' cooling regime has been made possible by
implementing a regularization of the dynamics, in order to get rid of
the inelastic collapse problem. Inelastic collapse~\cite{mcnamara} is
a singularity of the inelastic hard spheres dynamics which is
triggered by rare collisions happening at very low relative velocities
and in situations of high density, and it is not related to the shear
and clustering instability. Many kinds of regularization have been
proposed: in the rest of our paper we will use the one implemented by
Ben-Naim et al.~\cite{bennaim}, the so-called elastic cut-off: all
collisions happening at a relative velocity $|(\bv_1-\bv_2)\cdot
\bs|<v_e$ are treated as elastic. We (as well as in Ben-Naim's work)
have verified that the statistics behavior (e.g. $T_g$ and $\tau$
vs. $t$) does not depend on the choice of $v_e$, provided that $v_e
\ll \sqrt{T_g(t)}$. The percentage of those ``fake'' elastic
collisions remains negligible, but they are enough to prevent the
collapse.

The main characterization of the asymptotic regime, independently of
the regularization mechanism, is the granular temperature decay, $T_g
\sim t^{-1}$ in $2$ dimensions. The $t^{-1}$ law could be generally
explained as the consequence of some kind of diffusion, where an
energy decay $t^{-D/2}$ is expected. While this seems true for the
velocity field at the beginning of the instability, i.e. still in the
presence of a homogeneous density distribution~\cite{bettolo}, it
becomes less evident in the asymptotic, strongly compressible,
situation. Ben-Naim and coworkers have conjectured the belonging of
this ``final'' cooling stage to the universality class of the Burgers
equation: this is reasonable for a 1D inelastic gas, which is similar
to a gas of sticky particles, but it has proven to be incorrect in
higher dimensions~\cite{alain}. A mode-coupling theory, put forward by
Brito and Ernst~\cite{brito}, suggests an asymptotic decay of $T_g
\sim \tau^{-D/2}$. While in many simulations (see for example the
inset of right frame in Fig.~\ref{fig:global}) the asymptotic
relation between $\tau$ and $t$ seems to be linear, this is not true
for any value of $\alpha$. On the contrary. the asymptotic temperature
decay does not depend upon $\alpha$. Isobe~\cite{isobe} has shown the
remarkable (and unexplained) similitude between the final cooling
regime and two dimensional turbulence in incompressible fluids. Miller
and Luding~\cite{luding_new}, after having verified that the exponent
of the algebraic decay of $T_g(t)$ is $\sim -1$ also in $D=3$ have
suggested that the formation of clusters is a sort of percolation
process. Meerson and Puglisi~\cite{meerson} have shown that in $D=2$
elongated geometries (such that one direction is stable and the other
is unstable) the HCS is followed by an intermediated {\em
flow-by-inertia} regime, and finally a coalescence of quasi-singular
clusters {\`a la} Burgers.  An observation that we report here
(reserving details in further communications) is the strong narrowing
of scale separation between microscopic and macroscopic dynamics,
which makes difficult in practice a hydrodynamic description of this
final cooling regime~\cite{meerson2}.

\section{The network of collisions and the k-cores analysis}

Let us introduce the $k$-core decomposition. Given a graph $G=(V,E)$ of $|V|=n$
vertices and $|E|=e$ edges, the definition from~\cite{Batagelj02} of $k$-cores
is the following

\begin{definition}\label{k-core} 
A subgraph $H = (C,E|C)$ induced by the set $C\subseteq V$
is a {\em $k$-core} or a core of order $k$ iff $\forall v \in C: {\tt
  degree}_H(v)\geq k$, and H is the maximum subgraph with this property.
\end{definition} 

A $k$-core of $G$ can therefore be obtained by recursively removing
all the vertices of degree less than $k$, until all vertices in the
remaining graph have degree at least $k$. It is worth remarking that
this process is not equivalent to prune vertices of a certain
degree. Indeed, a star-like subgraph with node with a high degree that
connect many vertices with degree one and connected only with a single
edge to the rest of the graph is going to belong only to the first core
no matter how high is the degree of the node. We will also
use the following definitions

\begin{definition}\label{coreness} 
A vertex $i$ has {\em shell index} $k$ if it belongs to the 
$k$-core but not to $(k+1)$-core.
\end{definition} 

%
 
\begin{definition}\label{component}
A {\em component} is a connected subgraph, i.e. a path exists 
between any pair of nodes in the same component.
\end{definition}
 
\begin{figure}[htpb]
\includegraphics[width=4cm,angle=270]{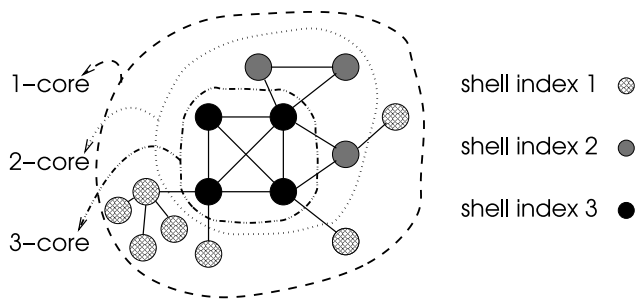}
\caption{\label{fig:kcores} This figure shows a graph and its
decomposition in $k$-cores. The {\em shell index} is the maximum core
that a node belongs to, then the $2$-core has nodes with shell index
$2$ and $3$.}
\end{figure}

The $k$-core decomposition therefore identifies progressively
internal cores and decomposes the network layer by
layer, revealing the structure of the different k-shells from the
outmost one to the more internal one (e.g. Fig.~\ref{fig:kcores}).

Now, we define a collision network as follows: each particle is a node
of the network, and each collision between two particles is
represented by a connection between both nodes. This network has the
particularity that each node may have multiples connections with other
nodes. Technically we say that a collision network is
not a graph, but a multigraph, i.e each edge may be multiple.

We have considered two possibilities to build a dynamical collision
network at time $t$. The first is obtained collecting all collisions
from time $0$ to time $t$: we call it a {\em complete} network. The
second one is built taking only the collisions occurred during a
window of time of fixed length (in terms of number of collisions),
ending at time $t$: this has been called a {\em partial}
network. Qualitatively the analysis of both networks produce the same
results: the distribution of populated shells is narrow in the
uncorrelated/homogeneous phase and is very spread in the
correlated/inhomogeneous one. The shell distribution in the partial
network has the advantage of being concentrated in a fixed range of
shells indexes, while the complete network distribution shifts with
time toward high shell numbers, since the average number of links per
node is growing. On the other side, the complete network offers a
better resolution of higher cores, which becomes crucial in the
non-homogeneous regime: during this regime, defining a characteristic
time scale and choosing the length of the time window for the partial
network gets harder and harder.  For this reason, in the following, we
have focused our attention on the complete network.


It is important to remark that the time complexity to compute the
$k$-core decomposition is very low. The vertex degree $d$ is given by the
number of nearest neighbors. Given a graph $G$ represented by its list
of vertices, where each vertex has a list of neighbors, the $k$-core
decomposition can be computed as following:
\begin{itemize}
\item Make an ordered array of lists, where each list is composed by the
vertices with the same degree. This step takes ${\cal O}(n)$, where $n$ is the
number of vertices.
\item Compute each $k$-core, recursively, starting by the minimum degree
$d_{\min}$, until any vertex remains in the graph. The time complexity of this
step is ${\cal O}(e)$, where $e$ is the number of edges.
\end{itemize}
To compute each $k$-core, it is enough to cut the vertices, and their
corresponding edges, with degree lower than $k$. Note that when a node
loses a neighbor, its degree is decreased, therefore it changes its
degree-list and may possibly fall into a degree-list which is being
cut.  Then, the final time complexity for a general graph is ${\cal
O}(e+n)$, which is transformed into ${\cal O}(e)$ for a connected
graph (where $e> n-1$).

Using this algorithm, it is very easy to compute the
$k$-shell. Normally, the algorithm begins from the nodes in the
$k-1=d_{\min}$ list. All cut vertices form the $(k-1)$-shell, while the
surviving graph is the $k$-core. Then at each step, the $k$-core and
the $(k-1)$-shell are obtained.

It is important to observe that a $1$-core has no isolated vertices,
and a $2$-core has no trees. For example, given a tree $T$, computing
its $1$-core leads to the same tree $T$; if its $2$-core is computed,
the result is an empty graph. Only a graph with loops gives a non
empty $2$-core. In general, a graph containing a $n$-clique, i.e. a
graph with its $n$ vertices all connected among themselves, has 
a non empty $(n-1)$-core.

\section{Results: homogeneous systems.}
\label{sec:hom}

In this section we present results about the analysis of the collision
network in homogeneous systems. In particular we will analyze
the dynamical collision network  of four different models:

\begin{itemize}

\item
a ``small'' system of inelastic hard disks ($\alpha=0.9$) that cools
down staying in the HCS during its whole history;

\item
an elastic system ($\alpha=1$) that permanently remains homogeneous in
its equilibrium dynamics;

\item
a system of inelastic hard disks simulated with the so-called
homogeneous Direct Simulation Monte Carlo~\cite{dsmc} (DSMC). In the
homogeneous DSMC each particle can collide with any other particle
(spatial coordinates are discarded) and the random choice of colliding
couples is performed with a probability proportional to
$(\bv_i-\bv_j)\cdot \bs$, in order to reproduce the kernel
of the collisional integral in the Boltzmann equation;

\item
a pure randomly growing network where links (collisions) are added
with uniform probability on the set of all possible couples of nodes.

\end{itemize}

All the four considered systems have $10^6$ particles (i.e. vertices). In the
first case, the size of the system is compared to the critical size
given by linear stability analysis of the HCS.  This
analysis~\cite{ernst} shows that a minimum size $L_c=\frac{2\pi}{k^*}$
is needed to have unstable shear modes, with
$k^*=\frac{\sqrt{1-\alpha^2}\sqrt{3}}{2\lambda}$,
i.e. $L_c=\frac{4\pi\lambda}{\sqrt{1-\alpha^2}}\approx 29 \lambda$,
where $\lambda=L^2/(\sqrt{2\pi}N\sigma)$ is the mean free path. The
small system considered here has a size $200000 \times 200000$,
corresponding to a density $\rho=2.5 \times 10^{-5}$, an area fraction
$1.96 \times 10^{-5}$ and a mean free path $\lambda \approx 1.6 \times
10^4$, so that $L \sim 12.5 \lambda \ll L_c$.  The elastic system
instead has the size of the large inelastic system discussed in the
next section, i.e. $3163 \times 3163$ (it is smaller in units
of a diameter, but is much larger in units of a mean free path).

\begin{figure}[htbp]
\includegraphics[clip=true,width=8cm]{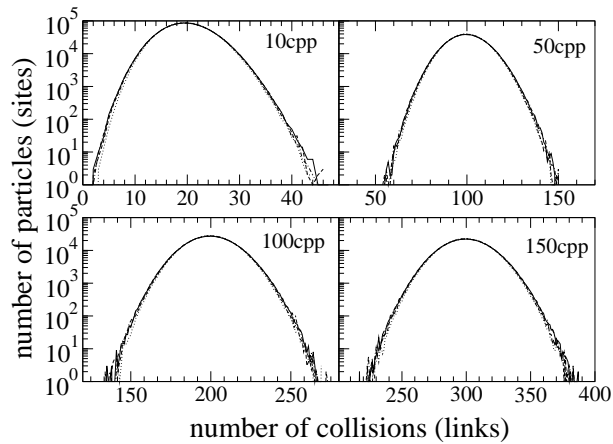}
\caption{\label{fig:deg_hom}Distribution of number of links per node
(collisions done by each particle) at different times for the four
homogeneous cases taken into account: a small inelastic system of hard
disks (solid line), a gas of elastic disks (dashed line), a gas of
inelastic disks simulated with the DSMC algorithm (dot-dashed line),
and a randomly growing network where each link/collision is done
choosing with uniform probability over the set of all possible pair of
nodes (dotted line).}
\end{figure}

In Fig.~\ref{fig:deg_hom} we present the evolution of degree
distribution $f(d,\tau)$ of $d$ edges attached to a node, with time
$\tau$ measured in collisions per particle ($cpp$), for the four
different cases. The average number of edges attached to each vertex
(i.e. of collisions done by one particle) is linearly growing with the
total number of collisions, i.e. $\langle d \rangle = 2 \tau$. The four systems display almost identical degree
distributions. The form of the distribution is close to a Poissonian
when $\langle d \rangle$ is small and becomes a Gaussian at large
$\langle d \rangle$. Poissonian and Gaussian statistics are the
natural expectations for homogeneous systems where no preferential
attachment is at work.

\begin{figure}[htbp]
\parbox{92mm}{
             \includegraphics[clip=true,angle=0,width=45mm]{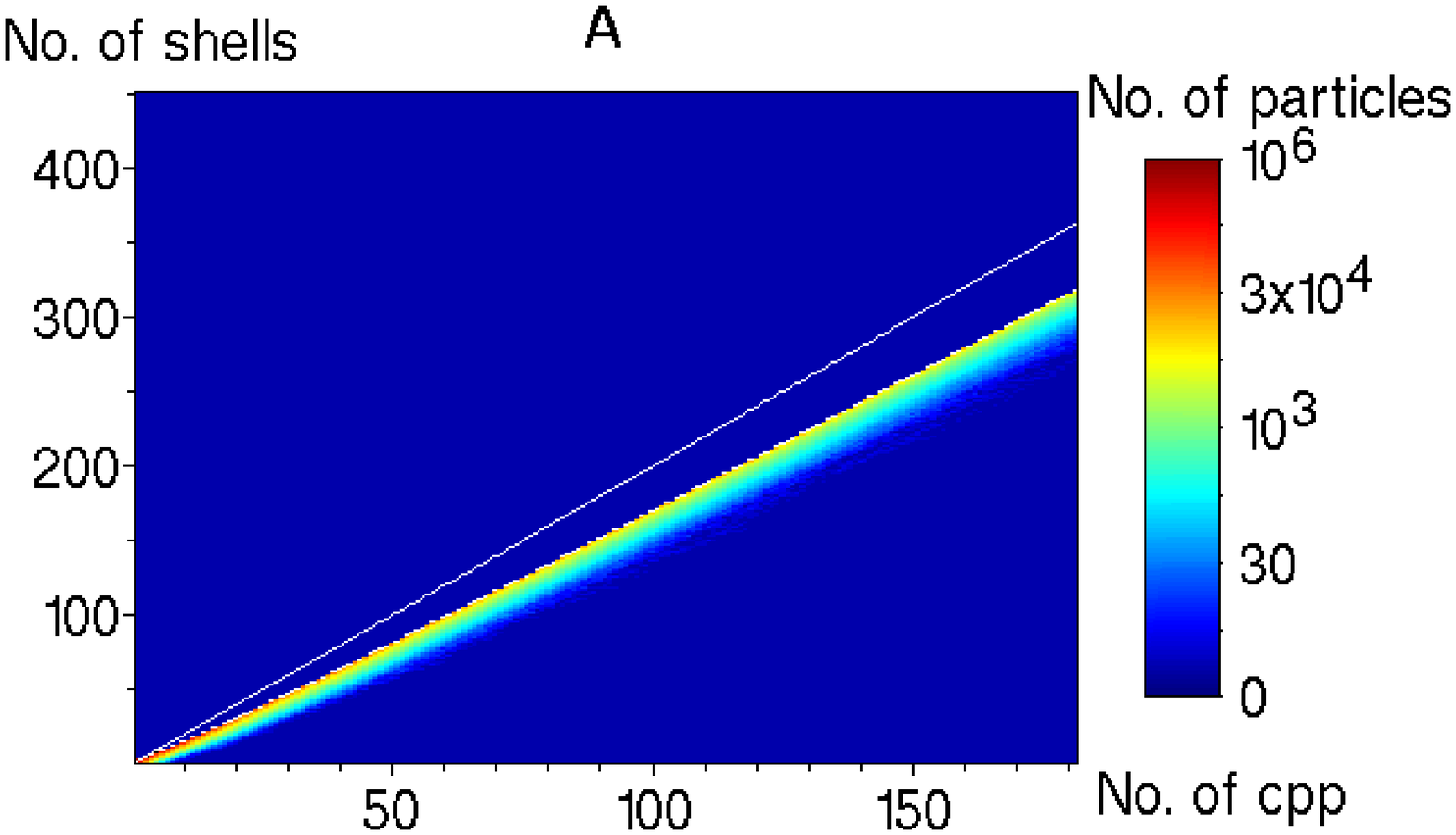}
             \includegraphics[clip=true,angle=0,width=45mm]{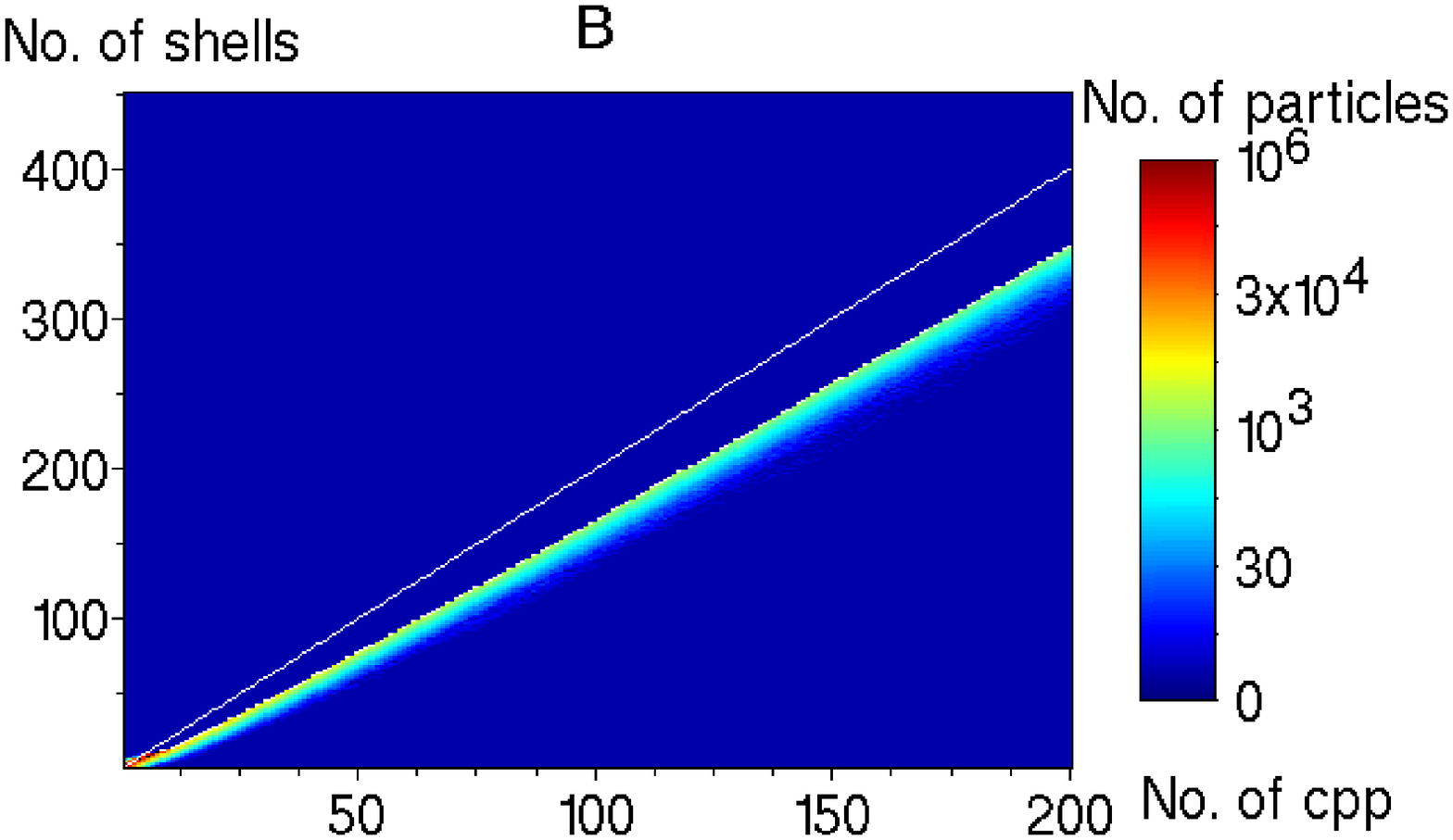}
             \includegraphics[clip=true,angle=0,width=45mm]{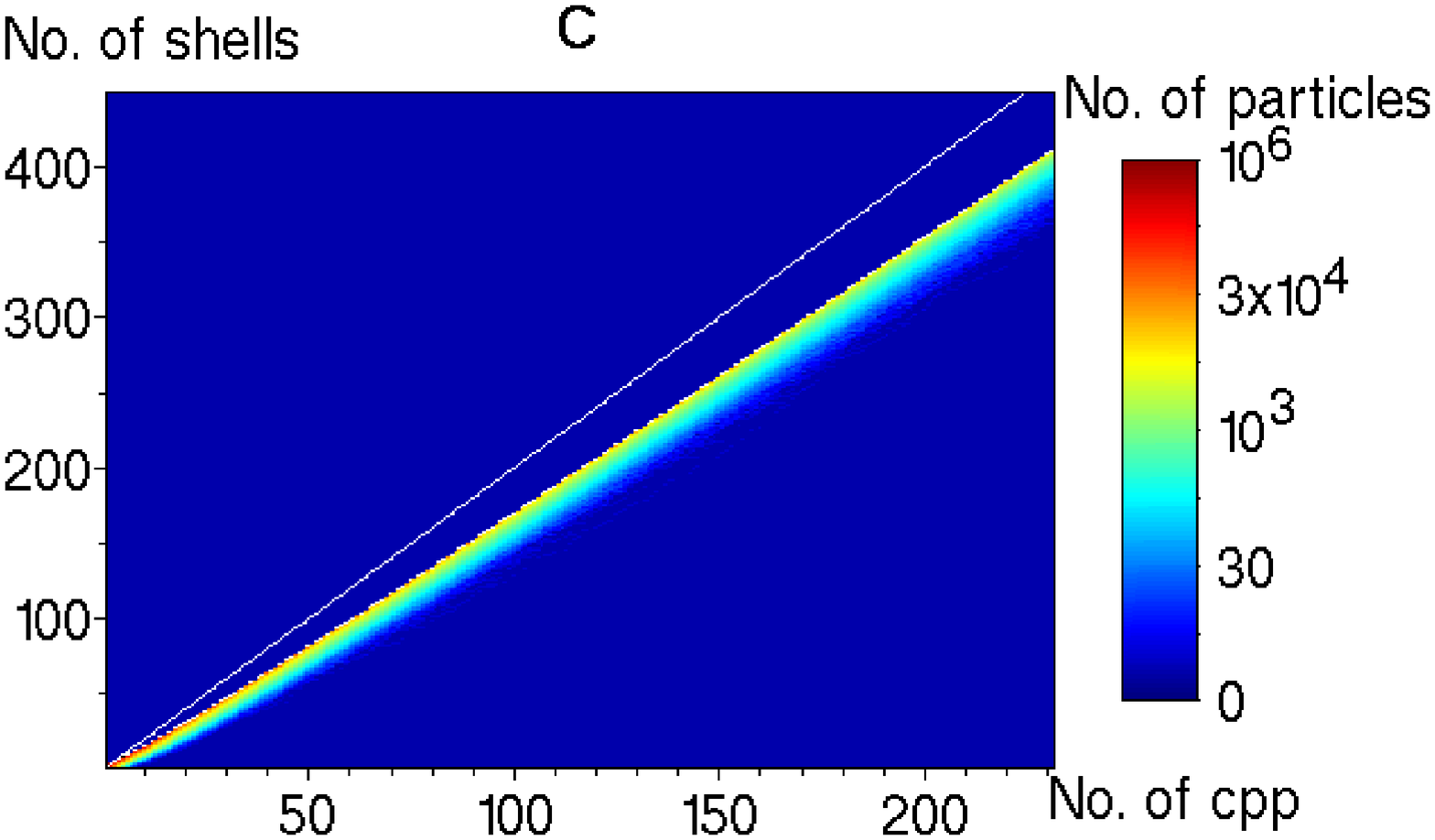}
             \includegraphics[clip=true,angle=0,width=45mm]{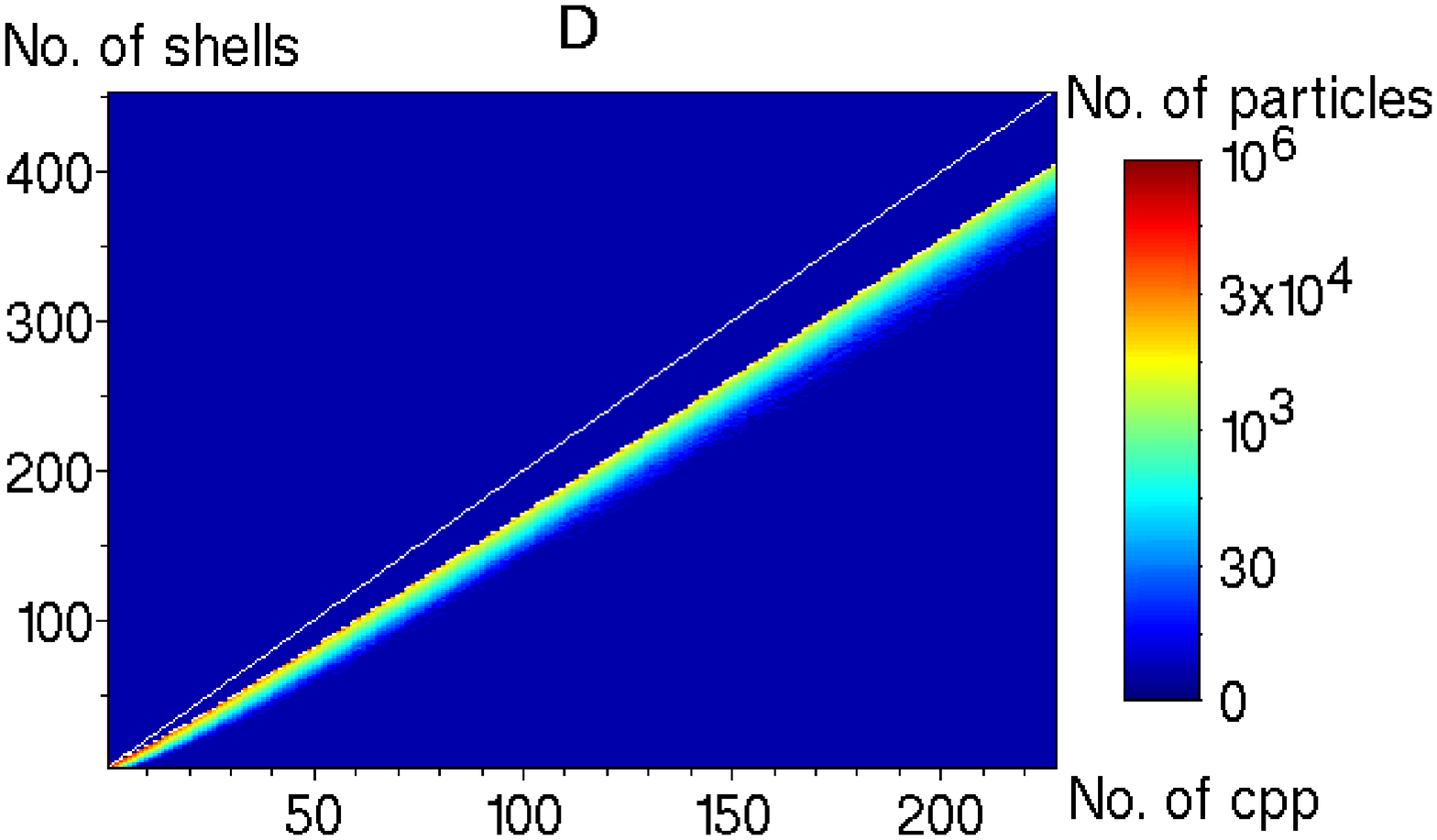}
}
\parbox{8cm}{\includegraphics[clip=true,width=8cm]{p_of_k_hom}}
\caption{\label{fig:kcore_hom}{\bf (Color online) }Distribution of
$k$-shells for the four homogeneous systems against time $\tau$
measured in collisions per particle (cpp). {\bf A-D}: time is on the
$x$-axis, while shell index $k$ is on the $y$-axis. Brightness/color
indicates population (number of particles in the shell). The four
sub-frames are: A) the small inelastic Molecular Dynamics (MD), B) the elastic MD, C) the
DSMC and D) the random network. {\bf E}: sections of the left graph at
four different times (solid line): the solid line is the small system
of inelastic hard disks, the dashed line is the elastic system of hard
disks, the (green) dot-dashed represents the system of inelastic hard
disks simulated with the DSMC algorithm, the dotted line is the random
network. The bold dashed (red) line in the two bottom graphs marks an
exponential growth $\sim exp(0.24 k)$.}
\end{figure}

In Fig.~\ref{fig:kcore_hom} results of the $k$-shell decomposition are
shown for the four different cases. Very similar results are observed
for all the cases: a small group of few shells is populated, the shape
of the distribution is stationary and is linearly shifting in time
towards high shell numbers. A shift in time is expected from the
simple fact that $\langle d \rangle$ is growing. Since the shape of
the distribution is conserved, we can track the growth of the average
shell number looking at the index of the maximally populated shell,
$k_{max}$. We observe for all the systems a growth $k_{max} \sim 1.8
\tau \sim 0.9 \langle d \rangle$. To our knowledge, the only
calculation of a similar quantity is in~\cite{doro}, in the case of
Erd\"os-R\'enyi graphs (i.e. random graphs with a Poisson degree
distribution with average $\langle d \rangle$) where the largest index
of a populated core grows as $0.78 \langle d \rangle$. Note that even our random
case is different, since we look to a random network with {\em
constant} total number of edges. On the other side, the shape of the
shell distribution seems almost identical in all the four homogeneous
case and strikingly resembles the shape of Erd\"os-R\'enyi graphs
depicted in Ref.~\cite{doro}, with an asymmetric distribution and
$k_{max}$ coinciding with the largest index of a populated shell. In
particular a nice exponential fit can be applied to the left tail of
the distribution with a growth rate that asymptotically (for large
$\tau$'s) does not depend upon $\tau$. 

A core of order $k$ can be, in principle, divided in components,
i.e. each of one is a connected subgraph and they are not connected
among them. We have verified that in all the four considered cases
only one component is present in the graph and in all subgraphs of any
core index, consistently with the scenario of the $k$-core
percolation: the maximum core is therefore a giant fully connected
component of the whole graph.

From the results of this section a first conclusion can be drawn on
the properties of a dynamical collision network: lack of correlation
in the process of link attachment gives place to a well defined,
narrowly peaked and shifting in time, distribution of shells. Even in
the presence of inelasticity we observe an almost perfect coincidence
with a randomly constructed network: this is not surprising, since the
only difference (the bias coming from the collision frequency due to
relative velocity) is not a preference on the degree of connection and
therefore does not affect the network topology.  Also the small
spatial correlations present in gases of hard disks with finite
diameter, do not play a relevant role: at the level of diluteness, we
had considered (area fraction smaller than $10 \%$), they give place
only to a small correction in the collision frequency, usually taken
into account with the so-called Enskog factor in the collisional
integral of the Boltzmann equation, with a small $\alpha$
dependence. In any case those finite diameter correlations do not bias
the attachment between nodes after a given total number of
collisions. Tiny discrepancies observed are, in our opinion, due to
finite statistics and are not relevant.

Before proceeding, the meaning of ``Molecular Chaos'' for real finite
systems can now be discussed. As mentioned in the introduction, the
only realization of a gas satisfying the true Molecular Chaos
constraint is a gas where each particle collides with particles that
are not connected (via other collisions) with its own previous
history: this of course requires infinite particles and gives place to
a collision network without cores of index larger than $1$, i.e. a
tree structure. In finite systems re-collisions are always possible
and they are reflected in the percolation of cores of growing indexes,
as shown in this section. Anyway, a vast literature teaches us that in
all systems discussed here, most of the predictions coming from
kinetic equations with the assumption of Molecular Chaos are
satisfied. As a matter of fact, the violations of Molecular Chaos
described by the presence of cores of growing indexes are negligible
for the most of physical observables. We propose, therefore, to use
in this case the term ``Finite Size Molecular Chaos''. This
assumption is analogous to the assumption, typically used in the
theory of random graphs, that a graph is ``locally a tree''.

A striking deviation from this behavior is observed when spatial
homogeneity is broken, as we discuss in the next section.

\section{Results: non-homogeneous cooling}

We consider here a large inelastic system breaking homogeneity
(specifically spatial translational invariance). Number of particles
is again $10^6$. System size is $3163 \times 3163$, the density is
$\rho=N/(L^2)=0.1$, corresponding to an area fraction of about $7.8
\%$ and a mean free path $\lambda \approx 4$: in this case $L \sim 790
\lambda \gg L_c$.  This system, after a first stage in the HCS,
develops strong and persistent structures in the velocity and
density field. Fig.~\ref{fig:global} clearly illustrates that after
$\sim 40$ cpp (collisions per particle) the HCS is no more a good
description of the time evolution of energy and collision
frequency. In particular, after $\sim 70$ cpp, the ``final'' regime is
reached where $T_g \sim t^{-1}$. As displayed in
Fig.~\ref{fig:frames}, during the whole ``final'' regime, a
continuous condensation of clusters is at work. Visual inspection of
the gas indicates the presence of groups of particles involved in a
huge number of collisions dissipating a very small fraction of the
energy, while the remnant particles constitute an even more dilute gas
and dissipate most of the energy in collisions among themselves or
against the clusters.

\begin{figure}[htbp]
\includegraphics[clip=true,width=8cm]{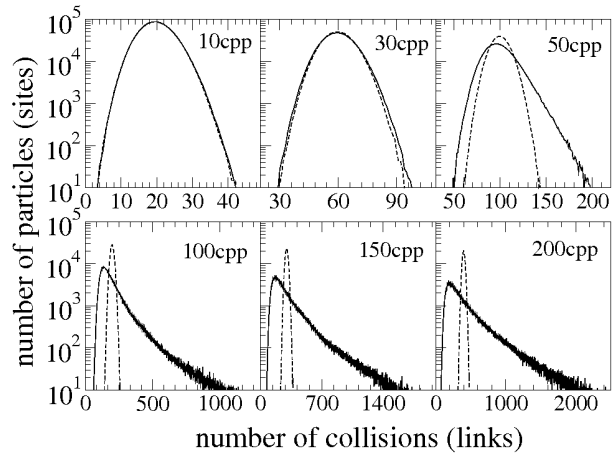}
\caption{\label{fig:deg_inhom}Histograms of degree distribution
$f(d,\tau)$ for different values of $\tau$, i.e. of number of
collisions per particle: comparison between the large inelastic system
(solid line) and the elastic system (dashed line). }
\end{figure}

The analysis of degree distribution of the dynamical network is given
in Fig.~\ref{fig:deg_inhom}, compared to the same measure for the
elastic system of the same size (which is representative of all
homogeneous systems discussed previously). Degree distribution starts
to deviate from the Gaussian behavior at $\sim 40$ cpp, i.e. together
with the departure from Haff law. The distribution develops a very
fat tail at large values of the number of links per node, representing
the contribution of particles trapped in clusters. No power law tails
are observed, but large stretched exponentials.

\begin{figure}[htbp]
\parbox{8cm}{\includegraphics[clip=true,width=8cm]{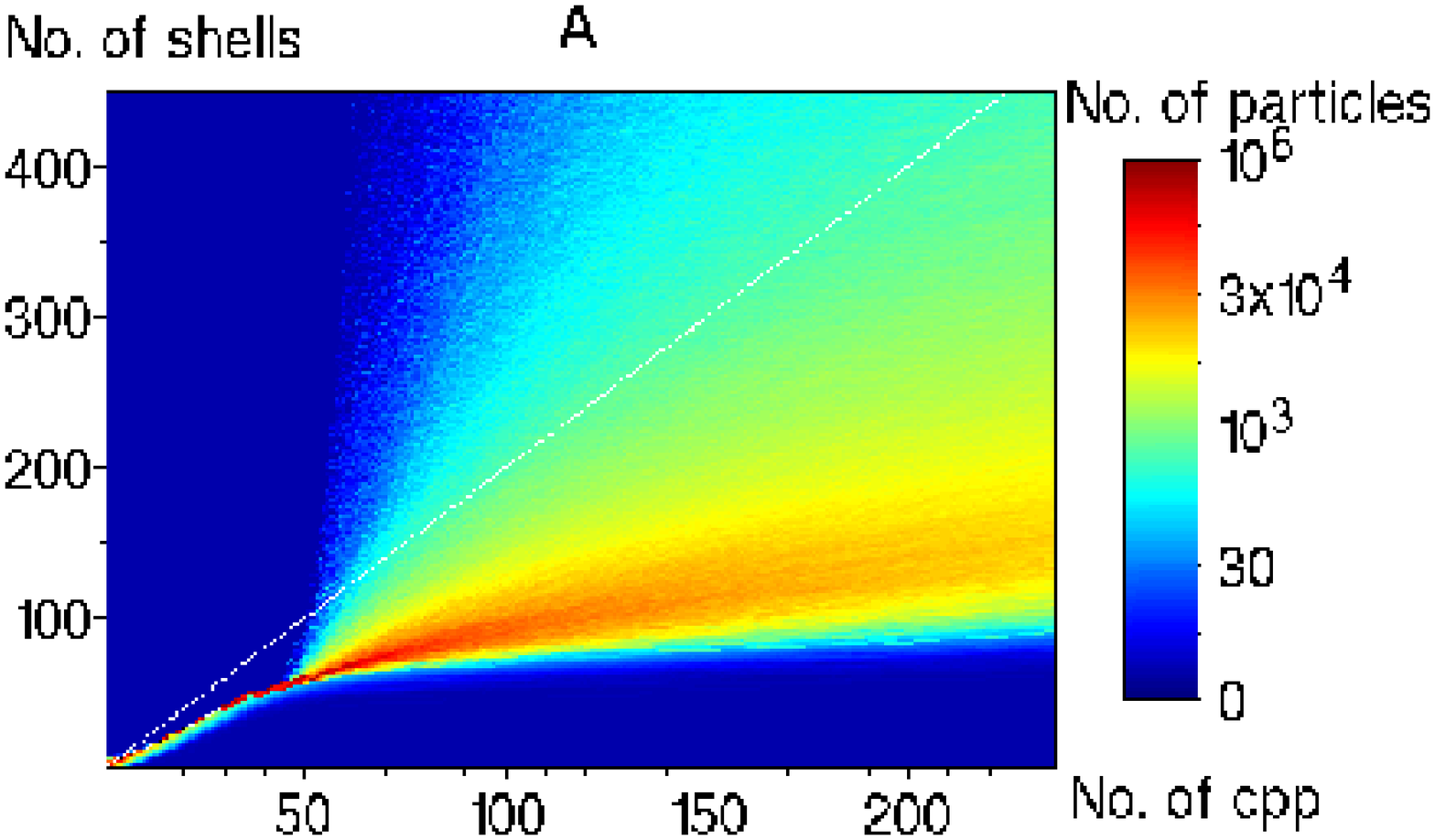}}
\parbox{8cm}{\includegraphics[clip=true,width=8cm]{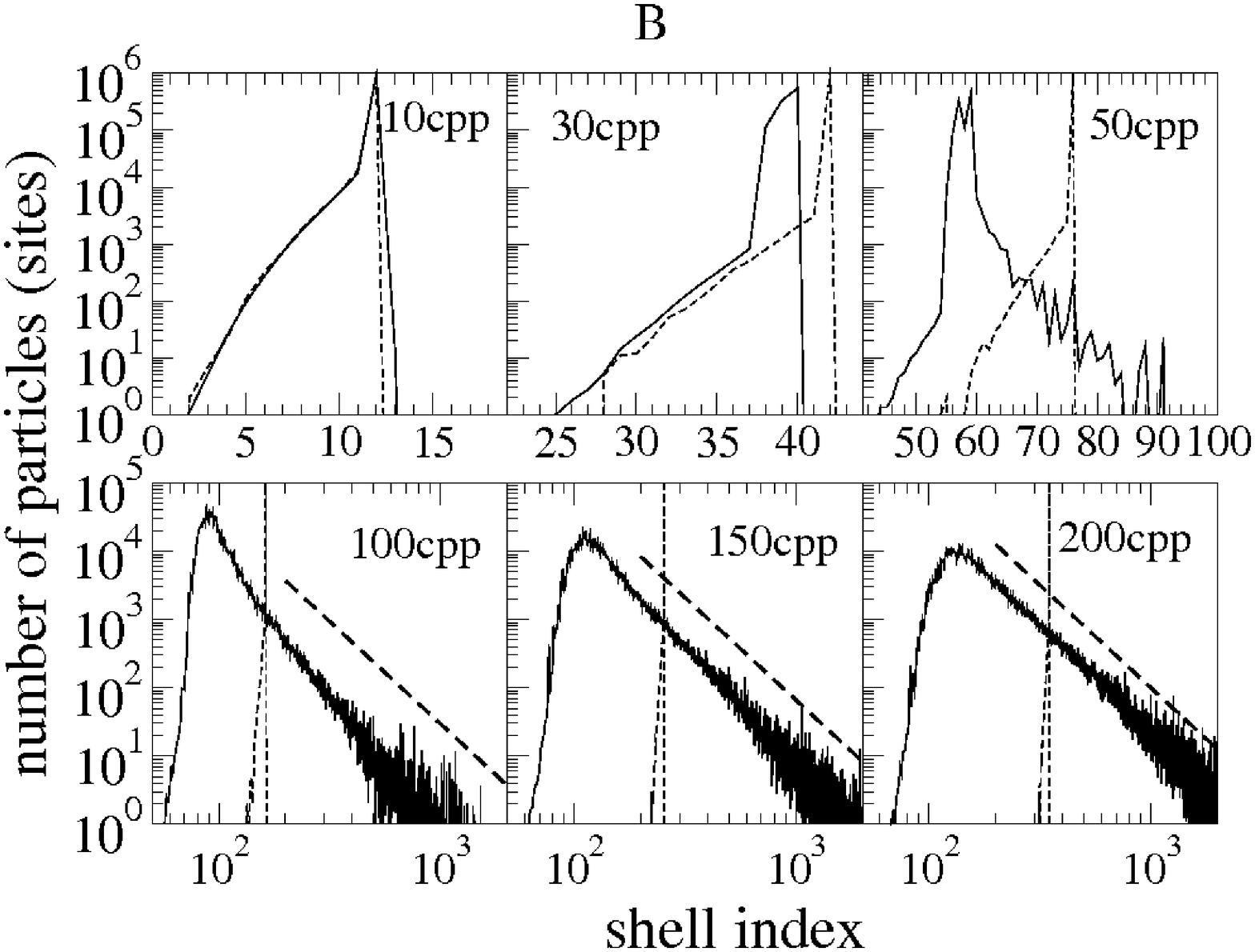}}
\caption{\label{fig:kcore_inhom}{\bf (Color online) } Distribution of
$k$-shells in the large inelastic system against time $\tau$ measured
in collisions per particle (cpp). {\bf A}: time is on the $x$-axis, while
shell index $k$ is on the $y$-axis. Brightness/color indicates
population (number of particles in the shell). The solid white line
marks the $1.8 \tau$ law observed in the homogeneous case. {\bf B}:
sections of the left graph at six different times (solid line),
compared with the elastic system of same size (light dashed
lines). The bold dashed line marks a power law decay with $-3$
exponent.}
\end{figure}

The $k$-core analysis of the dynamical collision network is presented,
giving results averaged over $10$ simulations starting from different
random initial conditions, in Fig.~\ref{fig:kcore_inhom}. Inspection
of frame A of this figure reveals an immediate and striking
difference with the corresponding plot for homogeneous systems in
Fig.~\ref{fig:kcore_hom}: at the time of deviation from the HCS
($\sim 40$ cpp) given by the departure from the Haff law, the narrow
distribution of populated $k$-shells starts to spread, while the index
of the maximally populated shell begins to deviate from the linear law
$k_{max} \sim 1.8 \tau$ discussed in the previous section, becoming
slower. In frame B, $k$-shell distributions at given times are
shown to better illustrate the phenomenon of very large cores
contagion. Shells with very high indices are populated by a consistent
fraction of particles. The tail at large $k$'s of the shell
distribution is well fitted by a power law decay with $\sim k^{-3}$,
while the small and medium $k$ range resemble a Gaussian function.

\begin{figure}[htbp]
\includegraphics[clip=true,width=8cm]{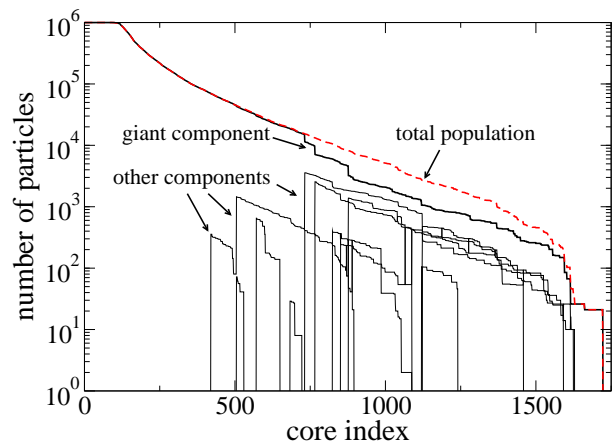}
\caption{\label{fig:200_comp}{\bf (Color online)} Plot of the number of particles in
disconnected components in each core, as a function of the core index,
at time $\tau=200$ cpp. At small core indexes, the number of observed
disconnected components is $1$, then it grows up to $\sim 6$.}
\end{figure}

The analysis of connected components in the whole graphs (core of
index $0$) and in the core subgraphs of higher index $k$, shown in
Fig.~\ref{fig:200_comp} for a given time, teach us that a large
component percolates the core, but at high indexes it is accompanied
by other few lesser components. A more detailed analysis reveals that
the different components are located in different spatial regions, as
expected: the disconnected components represent separate clusters.

\begin{figure}[htbp]
\parbox{8cm}{\includegraphics[clip=true,width=7cm]{200cpp_inhom}}
\parbox{8cm}{\includegraphics[clip=true,width=7cm,height=7cm]{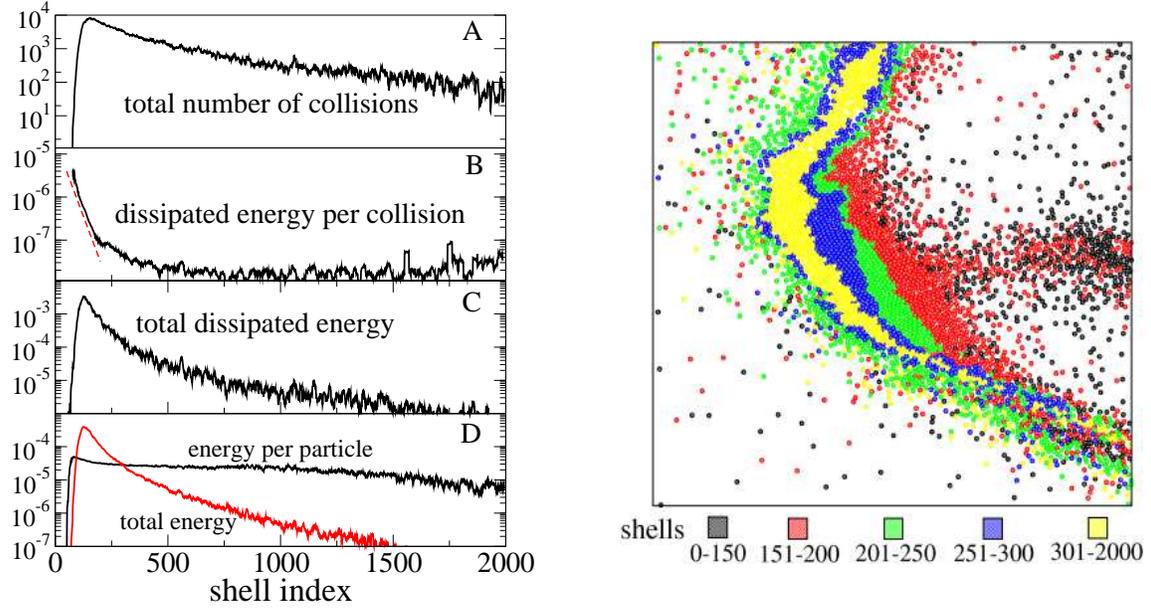}}
\caption{\label{fig:200_inhom}{\bf (Color online)} {\bf Left}:
statistics of physical observables in different shells at a given time
($\tau=200$ cpp). Units are arbitrary, depending on simulation
parameters. A) total number of collisions performed by particles in a
given shell. B) dissipated energy per collision, together with an
exponential fit of the first decay $\sim exp(-0.03 k)$. C) dissipated
energy summed over all collisions done by particles in a given
shell. D) average energy competing to each particle in a given shell
and total energy contained in a given shell, rescaled in order to
appear on the same scale. {\bf Right}: a snapshot of the system at
$\tau=200$ cpp, showing in different gray tones (colors) the shell
each grain belongs to. }
\end{figure}

Finally in Fig.~\ref{fig:200_inhom} frames A-D  we display some
relevant observables as a function of each shell at a given (large)
time $\tau$. The most remarkable observation concerns the average
dissipated energy per collision as a function of the shell index: a
strong decay (compatible with an exponential ramp) is visible. This
suggests that a typical scale $k^* \sim 100-150$ (choosing a
three orders of magnitude of difference), slightly depending upon
$\tau$, separates a group of shells where high dissipating collisions
occur against a larger group of shells dominated by weakly
dissipating collisions. This observation also provides a nice physical
interpretation of k-core decomposition: such decomposition fairly
manages to discriminate between the gas-like phase (few strongly
dissipating collisions) and the solid-like phase (many weakly
dissipating collisions).

Note that, while this separation is clear when the average dissipated
energy per collision is regarded, it becomes less evident when looking
at the total dissipated energy: collisions in high-index shells
dissipate few energy but dominate, in number, the total number of
collisions in the gas. We can still say that the most of the total
energy of the gas is dissipated in the first group of shells, but we
cannot put a clear discriminating mark between ``gas'' and ``solid''.

Another interesting observation concerns the kinetic energy contained
in the various shells. Total energy (i.e. summed over all particles in
a shell) still reproduces the large peak at small shell indexes
followed by a very large tail. On the other side the average kinetic
energy per particle has much weaker $k$-dependence: all shells contain
particles with similar kinetic energies, i.e. velocities in absolute
value. 

The spatial location of grains of a given shell index, given in the
right frame of Fig.~\ref{fig:200_inhom}, confirms the above physical
interpretation: particles in the first group of shells (indexes
$0$-$150$, at the time chosen) are spread in a low density/high
mobility phase, while the rest of the particles constitute compact
clusters: in particular, higher shell indexes individuate inner grains
whose escape probabilities are smaller and smaller.

\begin{figure}
 \begin{center}
 \includegraphics[width=171mm,angle=0]{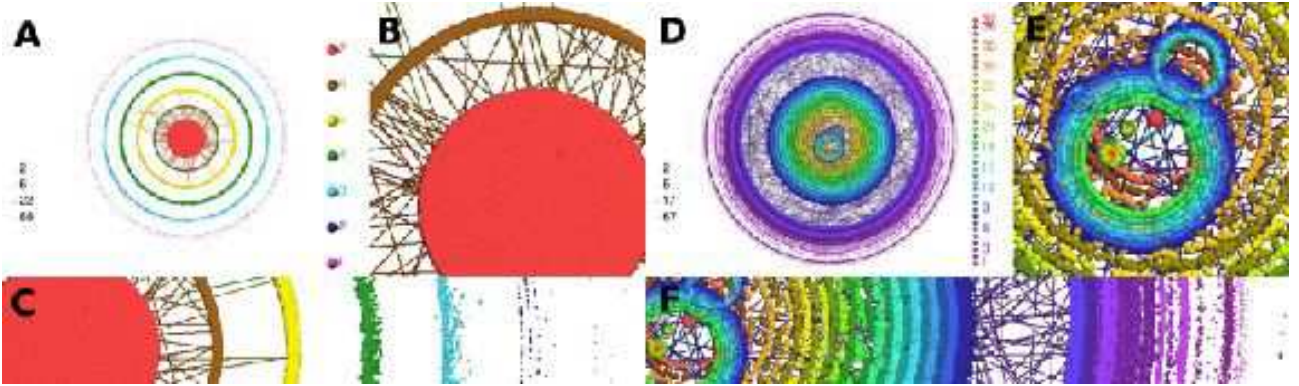}
 \end{center}
\caption{{\bf (Color online)} Visualization of the collision network
using LaNet-vi~\cite{LaNet-vi}. {\bf A-C}: an homogeneous casem, {\bf
D-F}: a non homogeneous case. Both visualizations have been done using
a uniformly sampled network because of computer memory constraints
(only $20\%$ of the collisions have been taken). Each sampled network
has the complete visualization (A and D), a zoom of a central part (B
and E), and an half of a longitudinal cut (C and
F).}\label{visualizations}
\end{figure}

Figure~\ref{visualizations} shows a synthetic representation of a part
of the total collision network after 100 cpp (i.e. $10^8$ links) for
both granular Molecular Dynamics (MD) simulations, i.e. homogeneous (A-C) and non
homogeneous (D-F). Only a fraction ($20\%$, i.e. $2 \times 10^7$) of
the total number of links is considered in the construction of the
picture: this leads to a distributions of shells that differs from the
previous analysis (core indexes are much smaller). All particles/nodes
are placed in the graph as dots: size, gray tone (color) and position reflect
various properties of that node. The gray tone (color) of a dot tells its shell
index (following the scale on the right). The size represents
the number of different particles linked to it (following the size
scale on the left). The position of each node is able to represent
both its core index and the connected component (in its own core) it
belongs: each core component is a group of circles with the same
center and radii proportional to the core index, while different
components have displaced centers, so that they can be
distinguished. Small (second order) modifications of this rule are
applied to represent also the number of links of the node toward
higher shells. In the visualization also a sample (only $1000$) of the
considered links is displayed, in order to give a flavour of the
density of connections in different parts of the network. Looking for
example at the right frame, one can clearly see that at low indexes
(violet) there is only one component, at high indexes (blue-cyan)
there are few different components, while at the highest indexes (red)
again only one component is present: all this is consistent with
Fig.~\ref{fig:200_comp}. The reader can find a complete description
in~\cite{nacho-viz} and\cite{LaNet-vi}. From this visualization it is
possible to immediately evaluate the difference between the
homogeneous and inhomogeneous case.  In the first case
(Fig~\ref{visualizations} frames A-C) there are very few shells and
most of the particles are in the maximum core, while all the cores
have a unique connected component, reminding a random graph
behavior~\cite{nacho}. On the contrary, in the non homogeneous case
(frames D-F of Fig~\ref{visualizations}), several components are
present, each representing a cluster, while the index of the maximum
core is much larger than in the homogeneous case.

\section{A model for recollisions}

Here we propose a minimal algorithm to stochastically build up a network
with properties similar to those observed in the collision network for
homogeneous and non-homogeneous granular gases. The number of nodes,
$N$, is constant with time, while links connecting pair of nodes are
added in sequence. A parameter $p_{hom} \in [0,1]$ denotes the
percentage of links homogeneously created. The algorithm consists in
sequential steps, at each step the following happens:

\begin{enumerate}

\item the first node $i$ is chosen with uniform probability among the $N$ nodes;

\item a random number $x$ is generated in the range $[0,1)$; 

\item {\bf if} $x<p_{hom}$ the second node $j$ is chosen with uniform
probability among the $N-1$ left nodes, i.e. $j \neq i$; a new link
$i-j$ is added; then repeat from 1;

\item {\bf if} $x\ge p_{hom}$, a sequence (``chain of recollisions'') of
$nr$ links is added; $nr$ is equal to the degree $d_i$ of node $i$
(number of links attached to $i$);

\begin{itemize} 

\item {\bf for} 1 {\bf to} $nr$ {\bf do}

\begin{itemize}

\item choose uniformely one of $i$'s links, and put another link to
      that neighbor

\end{itemize}

\item when the sequence of $nr$ ``recollisions'' ends, repeat from 1.

\end{itemize}

\end{enumerate}

\begin{figure}[htbp]
\parbox{8cm}{\includegraphics[clip=true,width=7cm]{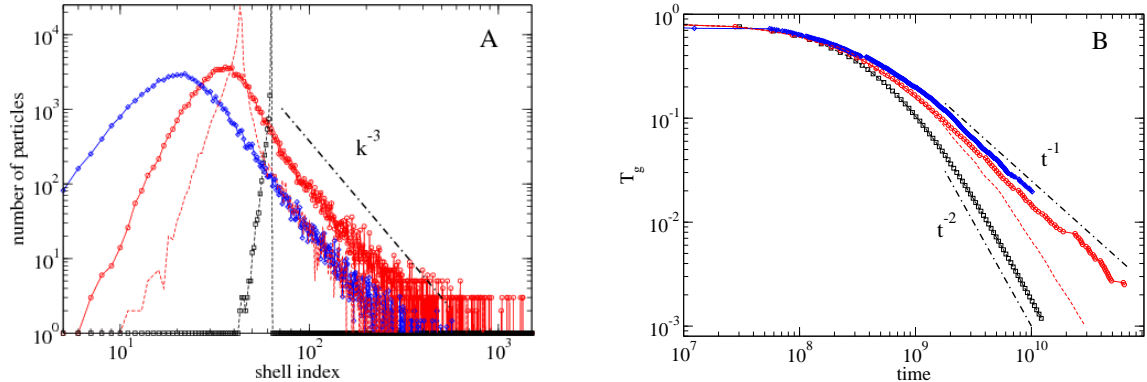}}
\parbox{8cm}{\includegraphics[clip=true,width=7cm]{model_ene}}
\caption{\label{fig:model}{\bf (Color online)} Simulations of the
recollision model. A) The number of nodes per shell, with $p_{hom}=1$
(black dashed curve with squares) after that $40N$ links have been
added, $p_{hom}=0.9$ (red solid curve with circles) after that $40N$
links have been added, $p_{hom}=0.8$ (blue solid curve with diamonds)
after that $25N$ links have been added and a case (red dashed line)
with $p_{hom}=0.9$ and $nr$ chosen randomly uniformly in $[1,d_i]$,
after $40N$ links. In all cases $N=10^5$. The thick dot-dashed line is
a $k^{-3}$ algebraic decay. B) The decay of the energy in the
recollision model, for the same cases. The two dot-dashed lines denote
the two algebraic decays $t^{-2}$ and $t^{-1}$. Here $N=10^6$ and
$\alpha=0.9$. Units are arbitrary, depending on simulation
parameters.}
\end{figure}

The results of the simulations of this simple model appear in
Fig.~\ref{fig:model}A. When $p_{hom}=1$, the results of all
homogeneous models are reproduced. The distribution of shells is
peaked at a value $\sim 0.88 \langle d \rangle$ and all populated
cores are spanned by a single connected component. When $p_{hom}<1$,
the situation radically changes: the distribution of shells becomes
broad, fairly resembling those of Fig.~\ref{fig:kcore_inhom}, with
many components in the cores with high indexes. In particular the
Gaussian-like shape in the low-index range and $k^{-3}$ power law
decay for the population of shell with high indexes is
reproduced. These remarkable findings suggest that the above algorithm
contains the ingredients sufficient to explain the topology of
recollisions in non-homogeneous granular gases. We have also tempted
different algorithms, changing the kind of preferential bias: most of
them generated strongly different distributions. A variant of the
model where $nr$ is uniformly randomly chosen between $1$ and $d_i$
displays an intermediate behavior, with a peak at low-index shells,
characteristic of the homogeneous systems, and the $k^{-3}$ tail
observed in correlated systems. In summary, the mechanism depicted in
the rules 1-4 above seems the only one to fully recover the physical
distribution of shells and components.

To conclude this analysis, we now speculate about the possibility of
recovering the physical decay of the granular temperature using this
minimal model for recollisions. Assigning initial random velocity
vectors (in 2D) to all $N$ nodes, from a normal distribution, and
updating the velocities of linked nodes $i$ and $j$ by
Eq.~\ref{eq:rule} at every new link, we obtain a decay of granular
temperature. In the collision rule we randomly choose the angle (with
a fixed axis) of the unitary vector $\bs$ from a uniform probability
in $[0,2\pi)$. Our first observation concerns energy dissipation: as
in the granular gas (see Fig.~\ref{fig:200_inhom}B), lower shells
contain the most dissipating collisions. One can also study energy
dissipation in different kinds of collisions (those with $x<p_{hom}$
against those with $x \ge p_{hom}$): on average, taken over a large and
fixed number of total collisions, the energy dissipated in the first
kind is two order of magnitude larger than the second.

Finally, we have provided our network model with a physical time $t$:
at each new link/collision, time advances of a step
$dt=1/|(\bv_i-\bv_j)\cdot \bs|$. This allows us to measure the time
decay of temperature $T_g(t)$, which is displayed in
Fig.~\ref{fig:model}B. One can easily recognize the Haff decay $T_g
\sim t^{-2}$ in the homogeneous situation, $p_{hom}=1$, i.e. in the
absence of recollisions chains, and the non-homogeneous decay $t^{-1}$
when $p_{hom}<1$. The consistency of the $p_{hom}=1$ case with
the Haff law is expected since, for this choice of the parameter, the
model is equivalent to the DSMC model discussed as third case in
Section~\ref{sec:hom}: particles in fact collide at random, but their
collision rate is proportional $1/dt \equiv |(\bv_i-\bv_j)\cdot
\bs|$.  Note also that the variation with $nr$ random in
$[1,d_i]$ does not reproduce the $t^{-1}$ decay when $p_{hom}<1$. The
same happens, for example, if one takes $nr$ to be a fixed fraction of
$d_i$. Since the $k^{-3}$ tail is obtained also with those variations
of the model, we must conclude that the $t^{-1}$ energy decay is much
less related to such tail than to the shape of shell distribution of
low indexes. This is consisten with the observation that energy
dissipation is mostly concentrated in the low index range.

These are preliminar results that are being explored in an ongoing
investigation~\cite{ignacionew}. Nevertheless, already at this level,
they are striking and indicate that the proposed recollision mechanism
is close to the one at work in the asymptotic non-homogeneous state of
the granular gas.


\section{Conclusions}

In this work we have exploited a novel statistical method, originally
introduced in the theory of random graphs, to gather information about
recollisions (i.e. collisions among correlated particles) in a
cooling granular gas. We constructed a network using the collision
sequence up to a given time, then we decomposed it in the $k$-core and
$k$-shell structure. In the homogeneous state, the finiteness of the
system naturally induces a giant component that percolates the core of
index $k_{max} \sim 1.8 \tau$ where $\tau$ is time measured as a
number of collisions per particles. When the gas becomes
inhomogeneous, because of inelasticity and large size, the system
separates into two populations with dramatically different core
statistics: a large component percolates the low index $k$-cores,
while the rest of the gas (a non negligible part) is found in a finite
(small) number of disconnected components at very high index
$k$-cores, corresponding to spatial clusters. The dilute low-core
phase is characterized by strongly dissipating collisions, while in
the clustered high-core phase the dissipated energy per collision is
$\sim 100$ times smaller.


The whole phenomenology is well reproduced by a simple model which
distillates the crucial ingredients of the recollision mechanism. This model
consists in a randomly growing  network which, tuning a
``homogeneity'' parameter, displays the same features of both
homogeneous and non-homogeneous granular gases. The success of this
model is the ability to reproduce, when supplied with a reasonable
measure of physical time, the observed decay of temperature $T_g \sim
t^{-1}$ which has received many non-conclusive interpretations in the
previous literature~\cite{bennaim,luding_new,brito}. The lack of
spatial coordinates for particles, in the proposed algorithm, suggests
that the statistics of recollisions and the energy decay do not
directly depend on the spatial arrangement of clusters, which is
usually observed to happen in the form of string-like structures.

The collision network analysis can also be applied to any ``hard
core'' system, including systems with soft interactions that can be
simplified as piecewise hard-like potentials. A natural extension is
the study of hard disks and spheres gases at very high packing
fractions, with the hope of elucidating the glass transition in
off-lattice systems, as well as experimental studies of granular
materials, where the use of rapid cameras is today able to collect
huge amounts of collisional data.

\vspace{\baselineskip}

\noindent {\it Acknowledgments.--} The authors wish to thank the
Laboratoire de Physique Th\'eorique d'Orsay, at the University of
Paris-Sud, for hospitality and cpu time. J.I.A.H. acknowledges the
support of European Commission - contract 001907 (DELIS).
A. P. acknowledges the support of the European Commission grant
MERG-021847. The work of A. P. has been also supported by the EU under
RD contract IST-1940 (ECAgents).


\end{document}